\documentclass[doublecol]{epl2} 

\usepackage{amssymb}
\usepackage{amsfonts}
\usepackage{amsmath}
\usepackage{amsthm}
\usepackage{bbm}

\title{Time delay control of symmetry-breaking primary and secondary oscillation death}
\shorttitle{Time delay control of oscillation death} 

\author{A. Zakharova\inst{1} \and I. Schneider\inst{2} \and Y. N. Kyrychko\inst{3} \and K. B. Blyuss\inst{3} \and 
A. Koseska\inst{4} \and B. Fiedler\inst{2} \and E. Sch{\"o}ll\inst{1}}
\shortauthor{A. Zakharova \etal}

\institute{                    
  \inst{1} Institut f{\"u}r Theoretische Physik, Technische Universit\"at Berlin, Hardenbergstra\ss{}e 36, 10623 Berlin, Germany\\
  \inst{2} Fachbereich Mathematik und Informatik, Freie Universit\"at Berlin, Arnimallee 2-6, 14195 Berlin, Germany\\
  \inst{3} Department of Mathematics, University of Sussex, Falmer, Brighton, BN1 9QH, UK\\
  \inst{4} Systemische Zellbiologie, MPI f{\"u}r molekulare Physiologie, Otto-Hahn-Stra\ss{}e 11, 44227 Dortmund, Germany
}
\pacs{05.45.Xt}{Synchronization; coupled oscillators}
\pacs{02.30.Ks}{Delay and functional equations} 

\abstract{ We show that oscillation death as a specific type of oscillation suppression, which implies symmetry breaking, can be controlled by introducing time-delayed coupling. In particular, we demonstrate that time delay influences the stability of an inhomogeneous steady state, providing the opportunity to modulate the threshold for oscillation death. Additionally, we find a novel type of oscillation death representing a secondary bifurcation of an inhomogeneous steady state.}

\begin{document}

\maketitle

\section{Introduction}
Time-delayed couplings arise naturally in many types of networks, for instance in 
coupled lasers \cite{SOR13}, neural networks \cite{ROS05,MAS08,STE09}, electronic
 circuits\cite{RRE98}, or genetic oscillators \cite{TIA13}, due to finite signal transmission and processing times, and memory and latency effects. While investigating real-world systems, it is necessary to take time delay into account, since the presence of time delay is an inherent property of the vast majority of processes that occur in nature \cite{ERN09,ATA10}. Moreover, time-delayed coupling and feedback represent an important aspect of control \cite{SCH07}.
Previous theoretical and experimental works have shown that time delay can be treated as a control parameter and can stabilize initially unstable states. In particular, time-delayed feedback has been used to stabilize
unstable periodic orbits embedded in a deterministic chaotic attractor \cite{PYR92,PYR06a}, or generated by a Hopf bifurcation \cite{FIE07}, unstable steady states \cite{HOE05}, spatio-temporal patterns \cite{BAB02,BET04a,KYR09}, or control the coherence and timescales of stochastic motion \cite{JAN03}.
In coupled nonlinear systems and networks time-delayed couplings represent an ubiquitous feature \cite{SCH13} which can also be used to control stability.

Particular interest, besides the control of various synchronization patterns
in networks of oscillators \cite{PIK01,BOC06a,HAK06,SOR07,MOT10,FLU10b,WIL13,NKO13} 
has recently been paid to the suppression of oscillations through the coupling. There are two types of oscillation quenching known in the literature \cite{ZOU09a,PRA10,KOS10,KOS13}: amplitude death and oscillation death. The distinction between amplitude death and oscillation death is essential, since the underlying mechanisms are crucially different. Amplitude death appears as a result of the stabilization of an already existing steady state that is unstable in the absence of coupling. On the contrary, oscillation death occurs due to a newly created stable stationary state which breaks the symmetry of the homogeneous system. Therefore, amplitude death is represented by a symmetric homogeneous steady state, whereas oscillation death is characterized by an inhomogeneous steady state.
Oscillation death associated with the stabilization of inhomogeneous steady states has been shown to exist in various systems, e.g., tunnel diode circuits \cite{HEI10}, electrochemical \cite{CRO89}, chemical droplets \cite{TOI08} and biological systems like neuronal networks \cite{CUR10}, calcium oscillators \cite{TSA06}, genetic oscillators \cite{KOS10a}, stem cell differentiation \cite{SUZ11}. Oscillation death is especially relevant for biological systems, since it provides a mechanism for cellular differentiation. In contrast, amplitude death only involves the change of stability of homogeneous steady states.
Therefore, besides distinct underlying mechanisms, amplitude death and oscillation death are also different from the application viewpoint. Moreover, in contrast to amplitude death that has been extensively reported by many researchers, oscillation death is much less studied.

There are three main factors that can cause amplitude death: coupling through dissimilar variables \cite{KAR07}, mismatches between the frequencies of the oscillators \cite{MIR90}, and time delay in the coupling \cite{RRE98}. In contrast, oscillation death appears due to a different reason, i.e., the symmetry-breaking of the system: for instance in two coupled oscillators a uniform steady state splits into two
additional antisymmetric branches with nonzero amplitude. However, in spite of the distinct mechanisms, evidence for oscillation death in systems coupled through dissimilar variables has also been previously found \cite{LIU12c}. Recently, it has been shown that introducing frequency detuning allows for observing both amplitude death and oscillation death in one system. Moreover, for sufficiently large mismatch between the frequencies, there occurs a transition from amplitude death to oscillation death \cite{KOS13}. Latency times in the coupling, without or with additional propagation time delay, have been shown to lead to annihilation of amplitude and oscillation death \cite{ZOU13}, resulting in exactly the same linear stability analysis as in an equivalent single system with delayed feedback and latency \cite{HOE05}.

In the present study, we investigate oscillation death in a system of coupled Stuart-Landau oscillators, which represents a generic expansion of a nonlinear system near a supercritical Hopf bifurcation (HB). This system is representative for a large class of coupled nonlinear oscillators \cite{KAR07,ATA03,FIE09,CHO09,DHU09,KYR11,SCH13b}, whose collective dynamics is ubiquitous in various fields of physics, chemistry, biology, and technology.  We uncover a novel solution of secondary oscillation death manifested by a secondary bifurcation of inhomogeneous steady states from the branch of primary oscillation death. We demonstrate that under specific conditions the stability of the inhomogeneous steady state can be controlled by time delay in the coupling.  Moreover, we find that both non-symmetrical and anti-symmetrical solutions can be present in one system. Both simulation-based studies and analytical approaches are performed. Our findings are
very generally applicable, and thus should be relevant for many systems in nature and technology where the oscillations are either desirable for the proper functioning, or in contrast should be suppressed because stable stationary, symmetry-breaking states of operation are required.

\section{Model}
We consider the paradigmatic model of Stuart-Landau (SL) oscillators 
\begin{equation}
  \label{eq:model}
\dot{z} = f(z) = (\lambda+i\omega - \left|z\right|^2)z.
\end{equation}
where $z = r e^{i \phi}=x+iy \in \mathbbm{C}$, $\lambda, \omega \in \mathbbm{R}$. For $\lambda>0$, the uncoupled system exhibits self-sustained limit cycle oscillations with radius $r_0=\sqrt{\lambda}$ and frequency $\omega$.
In this work we focus on two delay-coupled identical Stuart-Landau oscillators 
{\small
\begin{equation} \label{SL_del}
\begin{split}
\dot{z}_1&=\left(\lambda +i \omega - | z_1 \left(t\right)|^2\right)\,  z_1 \left(t\right)+\varepsilon\left(x_2 \left(t-\tau\right) -x_1 \left(t\right)\right)\\
\dot{z}_2&=\left(\lambda +i \omega - | z_2 \left(t\right)|^2\right)\,  z_2 \left(t\right)+\varepsilon\left(x_1 \left(t-\tau\right) -x_2 \left(t\right)\right) ,
\end{split}
\end{equation} }
where $z_{1,2}=x_{1,2}+i y_{1,2}$ are complex variables, $\lambda>0$ is the bifurcation parameter, $\varepsilon>0$ is the coupling strength and $\tau>0$ is the time delay. Here, we assume that the coupling is real, diffusive and symmetrical. Consequently, the whole system is also symmetric, i.e., invariant, with respect to permutation of the two oscillators. However, the continuous rotational $S^1$ symmetry of the system is broken by the particular coupling we introduce. The system (\ref{SL_del}) has a trivial steady state $z_{1,2}\equiv 0$, which is always unstable. In the absence of coupling $(\varepsilon=0)$, both oscillators have solutions in the form of stable periodic orbits with frequency $\omega$.  

The concept of oscillation death refers to the appearance of a new nontrivial inhomogeneous steady state due to the coupling. In order to find such steady states
we transform to symmetric ($z_+$) and antisymmetric ($z_-$) variables:
\begin{equation}\label{symm_antisymm}
\begin{split}
z_+&=\tfrac{1}{2} \displaystyle\left(z_1+z_2\right)\\
z_-&=\tfrac{1}{2} \displaystyle\left(z_1-z_2\right).
\end{split}
\end{equation}
where $z_+$ and $z_-$ correspond to the symmetric ($z_-=0$) and antisymmetric ($z_+=0$) manifolds, respectively, in the complex ($z_1,z_2$) phase space. Eqs.~(\ref{symm_antisymm}) can be inverted as $z_1=z_+ + z_-$, $z_2=z_+ - z_-$.
The system in the new coordinates is given by:
\begin{equation}\label{sub}
\begin{split}
\dot{z}_+&=\tfrac{1}{2} \displaystyle\left(f\left(z_+ +z_-\right)+f\left(z_+-z_-\right)\right)\\
\dot{z}_-&=\tfrac{1}{2} \displaystyle\left(f\left(z_+ +z_-\right)-f\left(z_+-z_-\right)\right)-2\varepsilon Re z_- 
\end{split}
\end{equation}
Note that, due to the symmetry, both $z_+\equiv 0$ and $z_-\equiv 0$ are dynamically invariant subspaces.
Similar to \cite{FIE09}, in these subspaces the dynamical equations (\ref{sub}) can be simplified to be
\begin{equation}
\begin{split}
  \dot{z}_+&=\left(\lambda +i \omega - |z_+|^2\right) \, z_+\\
  \dot{z}_-&=0
\end{split}
\end{equation}
in the \begin{em}{symmetric subspace}\end{em} $Z_+=\{\left(z_+,z_-\right)~|~z_-\equiv0\}$, and
\begin{equation}
\begin{split}
 \qquad \qquad \quad \dot{z}_+ &= 0\\
  \dot{z}_- &= \left(\lambda +i \omega - |z_- |^2\right) \, z_- -2\varepsilon Re z_-  
\end{split}
\end{equation}
in the \begin{em}{antisymmetric subspace}\end{em} $Z_-=\{\left(z_+,z_-\right)~|~z_+\equiv0\}$.
In the symmetric subspace there exists a stable periodic orbit but no steady states besides the homogeneous one $\left(z_+,z_-\right)\equiv \left(0,0\right)$.
On the contrary, the antisymmetric subspace allows for the appearance of the new inhomogeneous solution branches $s1$ and $s2$, given by
$\left(z_+,z_-\right)\equiv\left(0,\pm z_{s1}\right)$ and $\left(z_+,z_-\right)\equiv\left(0,\pm z_{s2}\right)$, respectively.
Here $z_{sj}=x_{sj} + i  y_{sj}$, $j=1,2$ with
\begin{align*}
  x_{s1}&=\, \frac{-\varepsilon +\sqrt{\varepsilon^2-\omega^2}}{\varepsilon} y_{s1},  \\
  y_{s1}&=\, \sqrt{\frac{\lambda \varepsilon -\omega^2\ + \lambda \sqrt{\varepsilon^2-\omega^2}}{2 \varepsilon}};
\end{align*}
\begin{align*}
  x_{s2}&=\, \frac{\varepsilon +\sqrt{\varepsilon^2-\omega^2}}{\varepsilon} y_{s2},  \\
  y_{s2}&=\, \sqrt{\frac{\lambda \varepsilon -\omega^2\ - \lambda \sqrt{\varepsilon^2-\omega^2}}{2 \varepsilon}}.
\end{align*}

\section{Oscillation death without time delay}
\begin{figure}[]
\begin{center}
\includegraphics[width=0.49\textwidth]{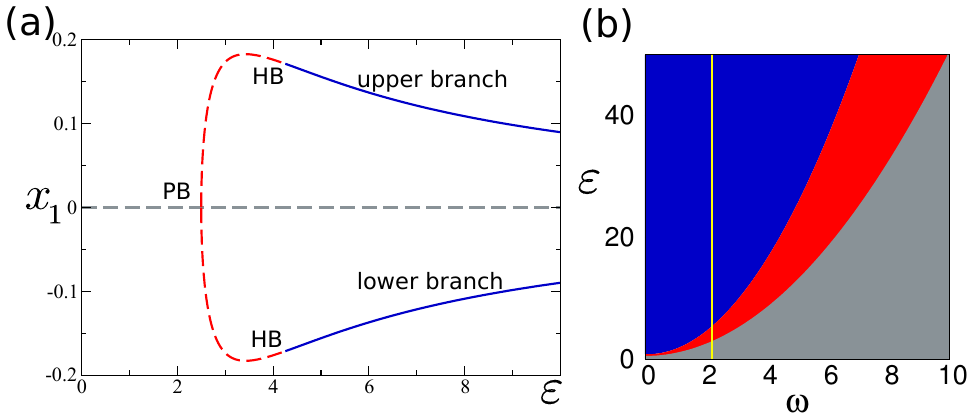}
\end{center}
\caption{(a) Bifurcation diagram showing the typical branches of the trivial homogeneous steady state (gray) and the inhomogeneous steady state (dark color) vs $\varepsilon$ for $\omega=2$. Solid lines denote stable steady states, dashed lines denote unstable steady states. Hopf bifurcation (HB) and pitchfork bifurcation (PB) are also marked. Numerically calculated using the continuation tool XPPAUT. (b) Regimes of inhomogeneous steady state in the parameter plane of the frequency $\omega$ and the coupling strength $\varepsilon$. Gray: inhomogeneous steady state does not exist, red (dark gray): inhomogeneous steady state exists but is unstable, blue (black): inhomogeneous steady state is stable. The vertical yellow (white) line indicates the value $\omega=2$ used in panel (a). Other parameters: $\lambda=1$, $\tau=0$.}
\label{fig:1}
\end{figure}

\begin{figure}[]
\begin{center}
\includegraphics[width=0.3\textwidth]{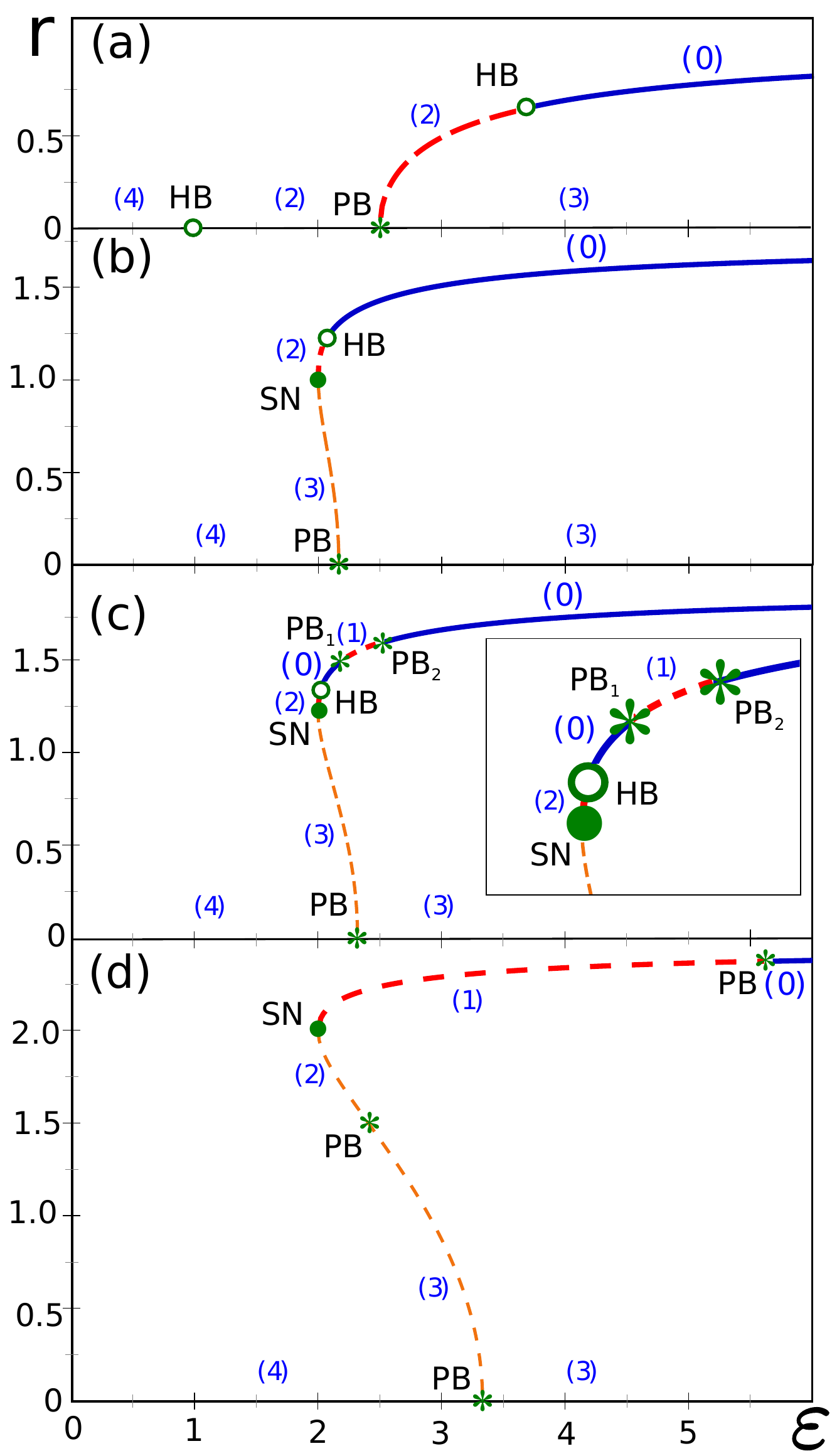}
\end{center}
\caption{Bifurcation diagrams for $\tau=0$, $\omega=2$ and different values of $\lambda$ (analytically calculated): (a) $\lambda=1$, (b) $\lambda=3$, (c) $\lambda=3.5$ and (d) $\lambda=6$. The numbers in parentheses (in blue) denote the unstable dimension of the steady state. $s1$-solution: thick (dark blue, red) lines, $s2$-solution: thin (orange) line. Empty circle: Hopf bifurcation (HB),  star: pitchfork bifurcation (PB,PB$_1$,PB$_2$), filled circle: saddle-node bifurcation (SN). Solid and dashed lines denote stable and unstable steady states, respectively.}
\label{fig:0}
\end{figure}

\begin{figure}[]
\begin{center}
\includegraphics[width=0.4\textwidth]{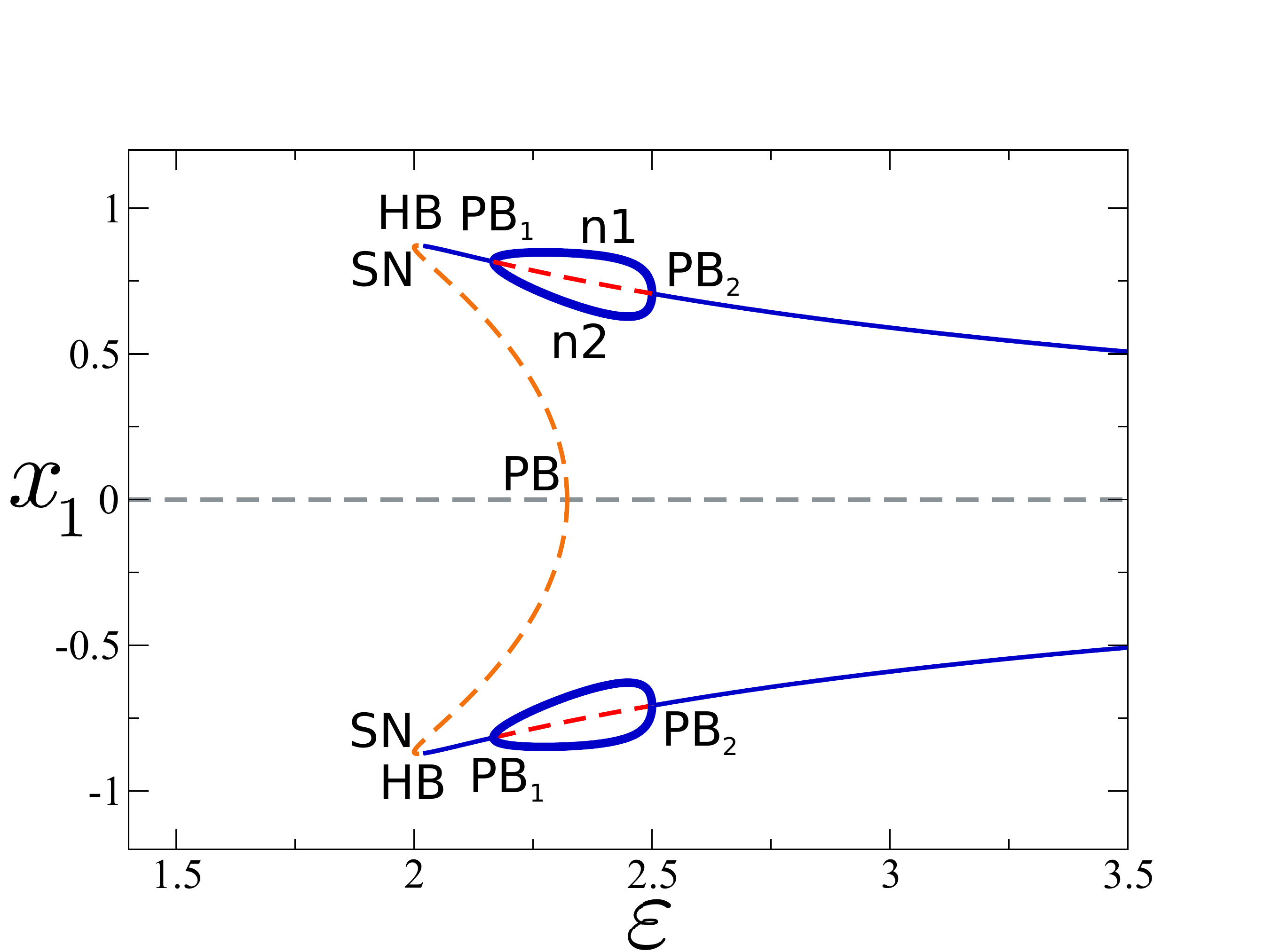}
\end{center}
\caption{Bifurcation diagram similar to Fig.~\ref{fig:1} but for $\lambda=3.5$, $\tau=0$, $\omega=2$. Secondary 
inhomogeneous steady state $n1$, $n2$ is indicated by thick closed blue curves. 
}
\label{fig:3n}
\end{figure}
As we shall see below, the solution branch $s2$ is always unstable whenever it exists.
Thus, generally, the inhomogeneous steady state of interest is given by $(x_1, y_1, x_2, y_2)= \pm\ (x_{s1}, y_{s1}, -x_{s1}, -y_{s1})$. This solution shows that for two coupled oscillators two different distributions of the oscillators between the branches of the inhomogeneous steady state are possible: (i) the first oscillator populates  the upper, whereas the second one populates the lower branch $(x_{s1}, y_{s1}, -x_{s1}, -y_{s1})$, or (ii) vice-versa, $(-x_{s1}, -y_{s1}, x_{s1}, y_{s1})$.

The existence of these inhomogeneous steady state branches depends upon the parameters $\varepsilon$, $\lambda$, and $\omega$. As an illustration, we consider
the case $\lambda < \omega$ in Fig.~\ref{fig:1}a, where $x_1$ is plotted versus the control parameter $\varepsilon$. Fig.~\ref{fig:1}b shows the different regimes of steady states in the ($\omega,\varepsilon$) parameter plane.
Fig.~\ref{fig:1}a demonstrates a typical bifurcation diagram for oscillation death at $\lambda=1$, $\omega=2$ and $\tau=0$. It corresponds to the yellow vertical line in Fig. \ref{fig:1}b at the fixed frequency $\omega=2$.
For small values of the coupling strength $\varepsilon$, the only steady state is an unstable homogeneous steady state (gray dashed line in Fig. \ref{fig:1}a and gray region in Fig. \ref{fig:1}b). At a critical value of the coupling strength $\varepsilon=2.5$ the symmetry-breaking occurs and a new solution representing the unstable inhomogeneous steady state (red dashed line in Fig. \ref{fig:1}a and red region in Fig. \ref{fig:1}b) is born via a pitchfork bifurcation (PB). The inhomogeneous steady state is manifested by two branches, upper and lower, which are stabilized via an inverse Hopf bifurcation at $\varepsilon=4.25$. Finally, the stable inhomogeneous steady state (dark blue solid line in Fig. \ref{fig:1}a and dark blue region in Fig. \ref{fig:1}b) indicates the regime of oscillation death. Moreover, this solution is preserved for different frequency values $\omega$ as shown in Fig. \ref{fig:1}b. 

For our analytical study \cite{footnote}, which we perform first in the case without delay, the system Eq.~(\ref{SL_del}) is converted to polar coordinates $z_-\left(t\right)=r\left(t\right)e^{i \phi \left(t\right)}$:
\begin{align}
  \dot{r}&=\left(\lambda -r^2 -2\varepsilon \cos^2\phi \right) \, r\\
  \dot{\phi}&=\omega +2\varepsilon \cos\phi\sin\phi
\end{align}
which gives the nontrivial steady state $s1$:
\begin{align}\label{rsteadystate}
  r&=\sqrt{\lambda - \varepsilon + \sqrt{\varepsilon^2-\omega^2}} \\
  \phi&=\arccos\left(\sqrt{\frac{\varepsilon+\sqrt{\varepsilon^2-\omega^2}}{2 \varepsilon}}\right)
\end{align}

Moreover, in order to provide general and precise theoretical conclusions on bifurcation scenarios which lead to oscillation death, we use here the notion of unstable dimension \cite{FIE09}. The unstable dimension of a steady state is the number of its eigenvalues with positive real parts. Let us consider first the simple case $\lambda<\omega$, for which except for the trivial homogeneous symmetric steady state there exists only the $s1$-solution. Fig. \ref{fig:0}a illustrates the bifurcation diagram of the radial coordinate $r$ in the antisymmetric subspace for $\lambda=1$ and $\omega=2$ and represents, therefore, an analytical counterpart of the numerical result shown in Fig. \ref{fig:1}a. The unstable dimension of the trivial steady state for $0<\varepsilon<\lambda$ is 4 (the unstable dimensions are indicated by brackets in the figure). At $\varepsilon=\lambda$ a 
Hopf bifurcation occurs in the symmetric manifold, and the unstable dimension of the trivial steady state changes from 4 to 2. Note that for $\lambda>\omega$ this bifurcation is not present.
For $\varepsilon=\frac{1}{2}\left(\lambda+\frac{\omega^2}{\lambda}\right)$, a pitchfork bifurcation takes place in the antisymmetric manifold, and the trivial steady state changes its stability again, from 2 to 3. The pitchfork bifurcation also gives birth to the inhomogeneous steady state with an unstable dimension 2, which is in exact agreement with the numerical result for $x_1$ in Fig. \ref{fig:1}a (red dashed line). 
For increasing $\varepsilon$, another Hopf bifurcation occurs in the antisymmetric manifold at $\varepsilon=\frac{1}{4}\left(\lambda+4\frac{\omega^2}{\lambda}\right)$ and the inhomogeneous steady state is stabilized, i.e, its unstable dimension becomes zero, which is also reflected in the numerical calculations (blue solid line in Fig. \ref{fig:1}a). Consequently, the strict analytical condition for oscillation death is that the unstable dimension of the inhomogeneous steady state becomes zero. For the case $\lambda<\omega$ this condition is fulfilled for $\varepsilon>\frac{1}{4} \left(\lambda+4\frac{\omega^2}{\lambda}\right)$.

For the range of parameters $\lambda>\omega$ 
(Fig. \ref{fig:0}b-d) there exist both the $s1$- and $s2$-solution, the latter exists for $\omega\le 
\varepsilon \le \frac{1}{2}\left(\lambda+\frac{\omega^2}{\lambda}\right)$ and is always unstable. 
The bifurcation scenario for the parameter values $\omega<\lambda<\sqrt{3}\omega$ is shown in Fig.~\ref{fig:0}b. Both $s1$- and $s2$-solutions are present and the $s2$-solution is unstable. The pitchfork bifurcation at $\varepsilon=\frac{1}{2}\left(\lambda+\frac{\omega^2}{\lambda}\right)$ gives rise to the antisymmetric inhomogeneous steady state, which appears on the $s2$-solution. There is a saddle-node bifurcation at $\varepsilon=\omega$ and a Hopf bifurcation at $\varepsilon=\frac{1}{4} \left(\lambda+4\frac{\omega^2}{\lambda}\right)$ on the $s1$-solution, which stabilizes the inhomogeneous steady state.

For $\sqrt{3}\omega<\lambda<2\omega$ 
(Fig. \ref{fig:0}c), we find an additional novel solution, which emerges from the $s1$-solution. It can be seen from the bifurcation diagram in Fig. \ref{fig:0}c that the inhomogeneous steady state has two stability regions (unstable dimension zero): one is between the Hopf bifurcation
at $\varepsilon=\frac{1}{4} \left(\lambda+4\frac{\omega^2}{\lambda}\right)$ and the pitchfork bifurcation $PB_{1}$ in Fig. \ref{fig:0}c at $\varepsilon=\frac{1}{3} \left(2\lambda-\sqrt{\lambda^2-3\omega^2}\right)$, and the other occurs after the second pitchfork bifurcation $PB_{2}$ in Fig. \ref{fig:0}c for $\varepsilon>\frac{1}{3} \left(2\lambda+\sqrt{\lambda^2-3\omega^2}\right)$. Consequently, between $PB_{1}$ and $PB_{2}$ the inhomogeneous steady state $z_{s1}$ becomes unstable. However, in this interval a secondary bifurcation branch $z_{n1}, z_{n2}$ appears (Fig. \ref{fig:3n}), which is also manifested by an inhomogeneous steady state. Therefore, for the choice of parameters $\sqrt{3}\omega<\lambda<2\omega$, we distinguish between primary and secondary inhomogeneous steady state. The secondary inhomogeneous steady state is a novel solution, which emerges from the $s1$-solution.
It is important to note that there are two types of oscillation death depending on the relation 
between the steady state values characterizing the subsystems: antisymmetric and non-symmetric. Previously, 
non-symmetric oscillation death has been shown to exist in chaotic oscillators without connection to secondary 
bifurcation \cite{LIU12c}, whereas in the present study we demonstrate non-symmetric oscillation death for a 
system of identical non-chaotic oscillators in a secondary bifurcation. For the two coupled Stuart-Landau 
oscillators the inhomogeneous steady state is antisymmetric if $x_1 = - x_2, y_1 = - y_2$ and non-symmetric if $x_1 \neq \pm x_2, y_1 \neq \pm y_2$. The primary inhomogeneous steady state is antisymmetric for all the parameter ranges considered here. In contrast, the secondary inhomogeneous steady state, which is observed for $\sqrt{3}\omega<\lambda<2\omega$ is of non-symmetric type:
it is given by $(x_1, y_1, x_2, y_2)= \pm\ (x_{n1}, y_{n1}, -x_{n2}, -y_{n2})$
or $(x_1, y_1, x_2, y_2)= \pm\ (x_{n2}, y_{n2}, -x_{n1}, -y_{n1})$. 
Therefore, for the model of two coupled identical Stuart-Landau oscillators we find both anti- and non-symmetric oscillation death regimes present in one system.  

In the case $\lambda>2\omega$ (Fig. \ref{fig:0}d) there also exist both $s1$- and $s2$-solutions. The regime of stable oscillation death occurs on the $s1$-solution after the inhomogeneous asymmetric steady state is stabilized via a pitchfork bifurcation at $\varepsilon=\frac{1}{3} \left(2\lambda+\sqrt{\lambda^2-3\omega^2}\right)$.

\section{Effect of time delay}
Next, we explore the effect of time delay upon oscillation death. In particular, we are interested in controlling the stability of the inhomogeneous steady state. 
By linearizing Eqs.(\ref{SL_del}) around the inhomogeneous steady states one can show that the $s2$-solution is always unstable. For the $s1$-solution we obtain a characteristic equation for the eigenvalue $\eta$, which can be factorized into two factors of the form
\begin{equation} \label{factors}
\eta^2+a \eta+ b_{1,2} \, \eta e^{-\eta \tau}+c+d_{1,2} \, e^{-\eta \tau}=0
\end{equation}
where the coefficients are given by
\begin{align*}
a&= 4 r^2-2 \lambda+\varepsilon\\
b_{1,2}&= \pm \varepsilon\\
c&= r^2\left(3 r^2-4 \lambda\right)+\lambda^2+\varepsilon \lambda-\varepsilon^2 + \left(\lambda+\varepsilon\right) \sqrt{\varepsilon^2-\omega^2}\\
d_{1,2}&= \pm \left(\varepsilon \lambda-\varepsilon^2-\omega^2 + \left(\lambda+\varepsilon\right)\sqrt{\varepsilon^2-\omega^2}\right)
\end{align*}
The steady state radius $r$ is given by Eq.~(\ref{rsteadystate}), hence the coefficients only depend on $\varepsilon$, $\lambda$, and $\omega$.
Delay neither affects the steady state nor zero-eigenvalue bifurcations. 
The stability of the steady states changes by Hopf bifurcation if two complex conjugate eigenvalues cross the imaginary axis.  
Setting $\eta=i \Omega$, $\Omega\neq 0$, Eq.~(\ref{factors}) becomes
\begin{equation}
0=-\Omega^2+i a \Omega+ i b_{1,2}\, \Omega e^{-i \Omega \tau}+c+d_{1,2}\,  e^{-i \Omega \tau}.
\end{equation}
Similar to \cite{FIE09}, we separate real and imaginary part of this equation.
Eliminating $\Omega$ from the resulting two equations gives a relation between $\tau$, 
$\varepsilon$, $\lambda$, and $\omega$, i.e., it defines the lines of Hopf bifurcations in the parameter space. By some lengthy algebraic manipulations \cite{footnote},
the following explicit form of the Hopf lines, $n=0,1,2,...$,
is obtained:
\begin{equation}\label{tau_eps}
\tau_n \left(\varepsilon\right) = \frac{1}{\Omega}\arccos \left(\pm\frac{\Omega^2\left(d-ab\right)-cd}{b^2 \,\Omega^2+d^2}\right) +\frac{2 \pi n}{\Omega} 
\end{equation}
with {\footnotesize
\begin{equation*}
\Omega^2=\tfrac{1}{2} \left( \left(-a^2+b^2+2 c\right) \pm \sqrt{\left(-a^2+b^2+2 c\right)^2-4 \left(c^2-d^2\right)}\right),
\end{equation*}}where $\Omega$, $a$, $b=b_1=-b_2$, $c$, and $d=d_1=-d_2$ depend only on $\varepsilon$, $\omega$, and $\lambda$, and $\pm$ corresponds to the first and second factor, respectively.
The stability boundary of the inhomogeneous steady state $s1$, i.e., the conditions for oscillation death, is a subset of these Hopf lines.

\begin{figure}
\includegraphics[width=0.49\textwidth]{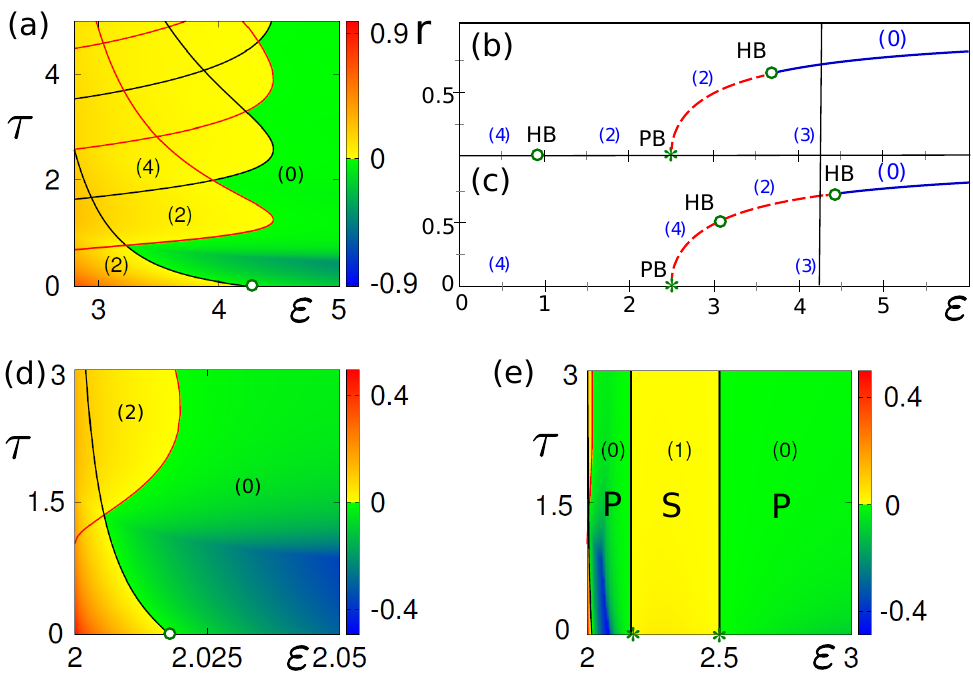}
\caption{(a) Modulation of stability regime of the inhomogeneous steady state by time-delayed coupling in the ($\varepsilon,\tau$) plane for $\omega=2$, $\lambda=1$ (analytical and numerical results). Black and red curves show the first and the second factor of the characteristic equation (\ref{factors}), respectively, analytically obtained from (\ref{tau_eps}). The numbers in parentheses denote the unstable dimension of the $s1$-solution. Color shaded regions indicate the maximum real part of the eigenvalues of (\ref{factors}) using {\it traceDDE} toolbox. (b), (c) Bifurcation diagrams for $\omega=2$, $\lambda=1$ and (b) $\tau=0.25$, (c) $\tau=1.25$. 
Empty circles denote Hopf bifurcations (HB), stars correspond to pitchfork bifurcations (PB). 
The $s1$-solution is shown by blue solid lines (stable) and red dashed lines (unstable). The 
vertical solid line marks the threshold coupling value for oscillation death ($\varepsilon=4.25$) in 
case of instantaneous coupling ($\tau=0$). (d), (e): Same as in panel (a) for $\omega=2$, 
$\lambda=3.5$. $P$ and $S$ denote primary and secondary oscillation death, respectively.} 
\label{fig:2} \end{figure}

Fig.~\ref{fig:2}a shows the results of the stability analysis for the 
case of $\lambda<\omega$. Both analytical $\tau_n\left(\varepsilon\right)$-curves and numerical calculations of the eigenvalues from the characteristic equation (\ref{factors}), computed using {\it traceDDE} toolbox, are combined in Fig. \ref{fig:2}a for $\lambda=1$ and $\omega=2$. The unstable dimensions following from the Hopf lines are given in brackets. The regime of stable steady state (0) coincides exactly with the domain where the largest real part of the numerically calculated eigenvalues is negative. By tuning the delay time $\tau$, the stability region of the inhomogeneous steady state can be either increased or decreased. Therefore, the threshold of oscillation death can be modulated by properly choosing the time delay. The bifurcations diagrams of $r$ vs $\varepsilon$ calculated analytically from (\ref{rsteadystate}) are shown in Fig. \ref{fig:2}b,c for two different values of the time delay. Oscillation death occurs, as in the case of instantaneous coupling, on the $s1$-
solution. Moreover, due to the presence of time delay, additional Hopf bifurcations may appear on the $s1$-solution and their number strongly depends on the value of time delay. For instance, for $\tau=1.25$ there is only one additional Hopf 
bifurcation at $\varepsilon=3.05$ (Fig. \ref{fig:2}c), whereas for $\tau=2.55$ their number is increased up to three (results not shown). Moreover, the  bifurcation diagrams clearly demonstrate the controllability of the inhomogeneous steady state by time delay. For instance, by choosing $\tau=0.25$ we observe oscillation death for a smaller value of the coupling $\varepsilon=3.67$ (Fig. \ref{fig:2}b) in comparison with the instantaneous case ($\varepsilon=4.25$), and $\tau=1.25$ shifts the threshold of oscillation death to a larger value of the coupling $\varepsilon=4.45$ (Fig.~\ref{fig:2}c).

For the parameter range $\omega<\lambda<\sqrt{3}\omega$, the structure of the stability regimes of the 
inhomogeneous state $s1$ is the same as in Fig. \ref{fig:2}a, and the Hopf bifurcation can again be shifted by 
delay, but the bifurcation diagram is similar to Fig.~\ref{fig:0}b with an additional unstable inhomogeneous steady 
state $s2$ appearing between the pitchfork (PB) and the saddle-node (SN) bifurcation. For $\lambda>\sqrt{3}\omega$ 
the modulation properties by time delay are much less pronounced. Fig.~\ref{fig:2}d shows for 
$\sqrt{3}\omega<\lambda<2\omega$ that the onset of primary oscillation death (P) at small $\varepsilon$ can still 
be tuned by time delay, but the subsequent onset of secondary oscillation death (S), where the antisymmetric inhomogeneous steady state has unstable dimension 1, is independent of delay, see the larger range of $\varepsilon$ plotted in Fig.~\ref{fig:2}e. This different behavior is due to distinct bifurcations: the controllable boundary of Hopf bifurcation, and the immutable boundaries characterized by zero eigenvalues $\eta$ of (\ref{factors}). The reason is that Hopf curves always change the unstable dimension by 2, and hence they cannot decrease the unstable dimension to 0. The same holds for $\lambda>2\omega$, corresponding to Fig.~\ref{fig:0}d.

\section{Conclusions}
We have considered oscillation death due to stable, but symmetry-breaking, steady states of two identical coupled oscillators. The stability boundary of Hopf type, where oscillation death occurs, is susceptible to efficient control by time delay. In particular, we 
have demonstrated that time delay strongly influences the stability of the antisymmetric 
inhomogeneous steady state, which is generated by a primary bifurcation of the symmetric homogeneous trivial steady 
state. Since time delay is ubiquitous in coupled systems in diverse areas, this provides a 
widely applicable mechanism to modulate the threshold for oscillation death. Additionally, we 
have found a novel type of non-symmetric oscillation death representing a secondary bifurcation of an 
inhomogeneous steady state. We have shown that the onset of this secondary oscillation 
death is not sensitive to time delay.

\acknowledgments
This work was supported by DFG in the framework of SFB 910.

\bibliographystyle{eplbib}

\end{document}